\documentclass[%
 reprint,
 superscriptaddress,
 nofootinbib,
 amsmath,amssymb,
 aps,
 pra,
]{revtex4-2}

\usepackage[justification=Justified]{caption}
\usepackage{threeparttable}
\usepackage{graphicx}
\usepackage{dcolumn}
\usepackage{multirow}
\usepackage[dvipsnames]{xcolor}
\usepackage{bm}
\usepackage[dvipsnames]{xcolor}
\usepackage[colorlinks=true]{hyperref}
\usepackage{natbib}
\usepackage{gensymb}
\usepackage{lineno}
\usepackage{mathtools}
\usepackage{bm,amssymb}
\usepackage{subcaption}
\usepackage{booktabs}
\usepackage{color} 
\usepackage{natbib}
\usepackage{physics}
\usepackage{siunitx}
\usepackage[dvipsnames]{xcolor}

\usepackage{comment}

\usepackage[colorlinks=true]{hyperref}
\hypersetup{
    linkcolor=blue,
    citecolor=blue,
    urlcolor=blue
}

\newcommand{\FQA}{Departament de Física Quàntica i Astrofísica, Universitat de Barcelona, 08028 Barcelona, Spain}
\newcommand{\ICCUB}{Institut de Ciències del Cosmos, Universitat de Barcelona, 08028 Barcelona, Spain}
\newcommand{\FIS}{Facultat de Física, Universitat de Barcelona, 08028 Barcelona, Spain}

\begin{document}

\preprint{APS/123-QED}

\title{Possible Solution to the Gallium Anomaly? Nuclear Transition Densities}

\author{D. Castillo}
\email{daniel.castillo@icc.ub.edu}
\affiliation{\FQA}
\affiliation{\ICCUB}
\affiliation{\FIS}

\author{D. Frycz}
\email{dorianfrycz@fqa.ub.edu}
\affiliation{\FQA}
\affiliation{\ICCUB}
\affiliation{\FIS}

\author{J. Menéndez}
\email{menendez@fqa.ub.edu}
\affiliation{\FQA}
\affiliation{\ICCUB}

\date{\today}

\begin{abstract}
The gallium anomaly, a $20\%$ deficit in the measured $\nu_e$ captures on $^{71}$Ga, 
remains unexplained. A recent work [Phys. Rev. Lett. {\bf 137}, 131805 (2026)] has proposed a solution based on the breakdown of the factorization of lepton and nuclear wavefunctions that underlie the prediction of the cross section, combined with ad hoc Gamow--Teller nuclear transition densities. We test this proposed solution with transition densities calculated within the nuclear shell model (NSM), normalized to the $^{71}$Ge electron-capture half-life and convolved with Dirac--Hartree--Fock--Slater electron wavefunctions. We explore several NSM interactions, bare and effective Gamow--Teller operators, as well as the impact of the valence space. In all cases, we obtain remarkably consistent NSM transition densities. Our factorized and nonfactorized cross sections agree, remaining $20\%$ higher than the measured value. In addition, we use NSM single-nucleon transition densities as ansatz and find that solving the anomaly requires fine-tuned densities, with percent-level cancellations and an overall scale orders of magnitude larger than the calculated NSM ones. We conclude that the gallium anomaly is not solved when using realistic nuclear transition densities.
\end{abstract}

\maketitle

\section{Introduction}
For over three decades, gallium experiments calibrated with intense radioactive sources have recorded fewer $\nu_e$ captures on $^{71}$Ga,
\begin{align}
\nu_e + {}^{71}\mathrm{Ga} \rightarrow {}^{71}\mathrm{Ge} + e^- \,,   
\end{align}
than theory predicts, a deficit known as the gallium anomaly~\cite{Elliott:2023cvh, Giunti:2010zu}. To verify their detectors' response to low-energy neutrinos, GALLEX~\cite{GALLEX:1992gcp, hampel_gallex_1999, Kaether:2010ag, GNO:2005bds} and SAGE~\cite{SAGE:2009eeu, doi:10.1142/9789811204296_0002} exposed their targets to $^{51}$Cr and $^{37}$Ar sources, and all four source runs produced less $^{71}$Ge than expected. Recently, the BEST experiment~\cite{Barinov:2021asz, Barinov:2022wfh} confirmed these results, measuring a deficit of about $20\%$ at a $\sim5\sigma$ level for two different Ga target volumes and a $^{51}$Cr source~\cite{Elliott:2023xkb}.

The most widely discussed explanation of the gallium anomaly involves physics beyond the standard model of particle physics, in particular short-baseline oscillations of $\nu_e$ into sterile neutrinos~\cite{GiuntiLaveder2007, GiuntiLasserre2019, Abazajian2012, Gariazzo2016, GonzalezGarcia2016, Diaz2020, Boser2020, DasguptaKopp2021, FarzanSchwetz2023, GiuntiTernes2024, Acero2024}. However, the parameter region favored by the gallium data has been progressively excluded. The KATRIN neutrino-mass measurement~\cite{KATRIN2025}, accelerator~\cite{MicroBooNE2025} and solar~\cite{Elliott:2023cvh, Goldhagen2022, GonzalezGarcia2025} neutrino data disfavor it, while very short-baseline reactor experiments~\cite{Choi2020, Danilov2021, Atif2022, Machikhiliyan2022, STEREO2023, PROSPECT2025} and combined analyses of reactor and gallium data~\cite{Giunti2022Gallium, Giunti2022Reactor} find no evidence for the corresponding $\nu_e$ disappearance~\cite{Elliott:2023cvh}. Very interestingly, there is no gallium anomaly when solar neutrinos are taken as source~\cite{Elliott:2026zsc}. With no consistent new-physics interpretation in sight~\cite{Huber2025}, attention has returned to the standard model prediction of the capture cross section.

Theoretical calculations of the cross section usually rely on the detailed balance approach~\cite{Bahcall1964, Bahcall1978, Bahcall1989, Bahcall1997}. In this scheme, one obtains the cross section for the $^{71}$Ga inverse $\beta$ decay (I$\beta$D) from the reversed process, the electron capture (EC) of $^{71}$Ge, assuming a common nuclear matrix element for both processes and computing the lepton part separately~\cite{Bahcall1989}. This requires that the nuclear and lepton contributions factorize. However, 
a recent study goes beyond the detailed-balance factorization scheme showing that, when combined with ad hoc fitted Gamow-Teller (GT) nuclear transition densities, it is possible to explain the gallium anomaly~\cite{Cadeddu:2025ueh}.

In this article, we follow on the latter work and implement the beyond-factorization scheme. Instead of using fitted nuclear transition densities, we obtain them in a many-body calculation, following previous $\beta$-decay and EC studies~\cite{Pastore:2017uwc,King:2020wmp,Ney:2021anu,Horiuchi:2021pup}. We use the nuclear shell model (NSM)~\cite{Caurier2005,Otsuka2020,Brown1988}, one of the leading many-body methods to investigate $\beta$-decay and EC, previously used to explore various aspects of the gallium anomaly~\cite{Kostensalo:2019vmv,Haxton:2025hye,Iwata:2025tzk}. We convolve the calculated NSM GT transition densities with electron wavefunctions computed using the Dirac--Hartree--Fock--Slater approach~\cite{salvat_radial_2019, Nitescu:2022tlh}. Within this framework, we assess whether it is possible to reproduce the $^{71}$Ge EC half-life and the $^{71}$Ga I$\beta$D cross section using the same NSM GT transition densities, thus solving the gallium anomaly.

\section{Inverse $\beta$ decay cross section}
The I$\beta$D cross section can be written as~\cite{Cadeddu:2025ueh},
\begin{equation}
    \sigma_{{\rm I\beta D}} = \frac{G^2_F|V_{ud}|^2 g^2_A}{\pi (2J_{\rm Ga}+1)}\sum_jB^jp_e^jE_e^j|\mathcal{H}^{j}_{\rm I\beta D}|^2 \mathcal{W}^j\,,
\end{equation}
where $G_F$ is the Fermi constant, $V_{ud}=0.9743$~\cite{navas_review_2024} the quark-mixing matrix element and $g_A=1.2764$~\cite{markisch_measurement_2019} the axial-vector coupling. The angular momentum of $^{71}$Ga is $J_{\rm Ga}=3/2$. The sum runs over atomic shells $j$ (neutrino lines), with branching ratios $B_j$, while $\mathcal{W}_j$ accounts for subleading contributions, weak magnetism and radiative corrections; both are taken from Table~I of Ref.~\cite{Elliott:2023xkb}. The momentum $p_e^j$ and total energy $E_e^j$ of the outgoing electron follow from energy conservation,
\begin{equation}
E_e^j = E_\nu^j - Q_{\rm EC} + m_e + \epsilon \,,
\end{equation}
where $m_e$ is the electron mass, $E_\nu^j$ the energy of the incident neutrino, $\epsilon=-0.09$~keV accounts for atomic rearrangement, and $Q_{\rm EC}=E_{\nu,1s}+B_{1s}$ is the $^{71}$Ge EC $Q$ value, related to the energy of the neutrino emitted in $1s$ capture, $E_{\nu,1s}=222.1$~keV, and the $K$-shell binding energy of Ga, $B_{1s}=10.37$~keV~\cite{cadeddu_reassessing_2026}. The transition amplitude reads, 
\begin{equation}
    \mathcal{H}_{\rm I\beta D}=\int 4\pi r^2\, \psi_e^j(r)\Psi^*_{\rm ^{71}Ge}(r)\hat{\mathcal{O}}_{\rm GT}\psi_\nu^j(r)\Psi_{\rm ^{71}Ga}({r})dr\,.
\end{equation}
Here, $\psi_e^j(r)$ and $\psi_\nu^j(r)$ represent the electron and neutrino wavefunctions, while $\Psi_{^{71}\rm Ga}(r)$ and $\Psi_{^{71}\rm Ge}(r)$ correspond to the initial and final nuclear states. The GT operator, $\hat{\mathcal{O}}_{\rm GT}=\sum_k \boldsymbol{\sigma}_k\tau_k^-$, acts in spin--isospin space and converts a neutron into a proton. Since we do not factorize the lepton and nuclear parts of the amplitude, lepton wavefunctions must be integrated together with the radial profile of the nuclear transition, encoded in the GT transition density,
\begin{equation}
C(r)=4\pi r^2 \Psi^*_{\rm ^{71}Ge}(r) \hat{\mathcal{O}}_{\rm GT}\Psi_{\rm ^{71}Ga}({r}) \,,
\end{equation}
with $\int_0^\infty C(r)\,dr=M_{\rm GT}$, the GT nuclear matrix element of the factorized detailed-balance scheme. 

We solve the Dirac--Hartree--Fock--Slater equations for the electron wavefunctions with the RADIAL package~\cite{salvat_radial_2019}, following Ref.~\cite{cadeddu_reassessing_2026}. We only keep the leading axial contributions for a GT transition~\cite{Cadeddu:2025ueh}, so that, 
\begin{equation}\label{eq:IBD}
    \begin{split}
        & |\mathcal{H}_{\rm I\beta D}|^2= |I_1|^2+|I_2|^2\,, \\
         I_1 = \int dr\, C(r)&\left[ g_{-1}(r)j_0({qr})+\frac{1}{3}f_{-1}(r)j_1({qr})\right]\,,\\
         I_2 = \int dr\, C(r)&\left[ f_{1}(r)j_0({qr})-\frac{1}{3}g_{1}(r)j_1({qr})\right]\,,
    \end{split}
\end{equation}
where $j_0(qr)$ and $j_1(qr)$ are spherical Bessel functions, which arise from the expansion of the neutrino plane wave with momentum $q=|\boldsymbol{q}|$. The continuum electron radial wavefunction is described by its large and small components, $g_\kappa(r)$ and $f_\kappa(r)$, where
$g_{-1}(r)$ and $f_1(r)$ [$f_{-1}(r)$ and $g_1$(r)] correspond to $s_{1/2}$ ($p_{1/2}$) partial waves. 

\section{Electron capture half-life}
We also compute the half-life of $^{71}$Ge, which decays by EC~\cite{Cadeddu:2025ueh, Bahcall1964, Bambynek:1977zz, Brysk:1958zz, Bahcall:1962zz},
\begin{equation}
    \frac{\ln 2}{t_{1/2}}=\frac{G^2_F|V_{ud}|^2 g_A^2}{\pi(2J_{\rm Ge}+1)}\,E_{\nu,1s}^2\,|\mathcal{H}^{\rm EC}_{1s}|^2\left[ 1+ \frac{P_L + P_M}{P_K} \right],
\end{equation}
where $J_{\rm Ge}=1/2$ is the $^{71}$Ge angular momentum, and $P_K$, $P_L$ and $P_M$ the experimental capture probabilities from the $K$, $L$ and $M$ shells~\cite{Bambynek:1977zz, EDELWEISS:2020fxc, collar_new_2023}, 
taking into account overlap-and-exchange corrections. The EC amplitude reads,
\begin{equation}
    \mathcal{H}_{1s}^{\rm EC} =\! \int_0^\infty dr\, C(r)\!\left[g_{1s}(r)\,j_0(qr)+\frac{1}{3}f_{1s}(r)\,j_1(qr)\right],
\end{equation}
where $g_{1s}(r)$ and $f_{1s}(r)$ are the large and small components of the bound $1s$ electron wavefunction, in contrast to the continuum wavefunctions that enter the I$\beta$D amplitude.

In order to reproduce the experimental weighted-average half-life, $t_{1/2}=11.465\pm 0.003$~d~\cite{cadeddu_reassessing_2026, collar_new_2023, Hampel1985, Norman2024, Derbin2025}, we introduce an overall quenching factor $q_{\rm EC}$ multiplying the nuclear transition density. We fix its value within the factorization scheme~\cite{Bahcall1964, Bahcall1978, Bahcall1989, Bahcall1997,Haxton:2025hye,Elliott:2023xkb},
\begin{equation}
    |\mathcal{H}_{1s}^{\rm EC}|^2 = \left[g_{1s}^2(R_{\rm nucl})+f_{1s}^2(R_{\rm nucl})\right] q_{\rm EC}^2\, M_{\rm GT}^2\, .
\end{equation}
Here $R_{\rm nucl}=\sqrt{5/3}\,R_{\rm ch}$ is the radius of the equivalent uniformly charged sphere, obtained from the measured $^{71}$Ge charge radius $R_{\rm ch}=4.032\pm0.002$~fm~\cite{wang_nuclear_2024}. Because we fit $q_{\rm EC}$ to the measured half-life, it effectively absorbs the impact of two-body currents~\cite{Menendez:2011qq,Gysbers2019, Stroberg2021,Li:2025exk}, weak-magnetism and radiative corrections~\cite{Elliott:2023xkb} to the EC rate, which we therefore do not include explicitly. While two-body currents and weak magnetism can modify the radial shape of the transition density---radiative corrections do not---these contributions are subleading, so we do not expect a sizeable effect. We then apply the same $q_{\rm EC}$ in the beyond-factorization calculation. 

\section{ NSM ${\rm GT}$ transition densities}
We obtain the NSM GT transition density from the one-body transition densities (OBTDs) between the $^{71}$Ga and $^{71}$Ge ground states~\cite{Suhonen2007},
\begin{equation}
C(r)=4\pi\,\sum_{ab}\mathrm{OBTD}_{ab}\,\hat{\mathcal{O}}_{\rm GT}^{ab}\,
R^{p}_{a}(r)\,R^{n}_{b}(r)\,r^{2}\,,
\end{equation}
with
\begin{equation}
\mathrm{OBTD}_{ab}\propto \langle ^{71}{\rm Ge}\|\,[c^\dagger_{a}\otimes\tilde{c}_{b}]^{(1)}\|^{71}{\rm Ga}\rangle\,,
\end{equation}
where $c^\dagger_{a}$ ($\tilde{c}_{b}$) creates (annihilates) a nucleon in the single-nucleon orbital $a$ ($b$), described by the principal quantum number $n$, the orbital and total angular momenta $l$ and $j$, and the projection $m_j$. $\hat{\mathcal{O}}_{\rm GT}^{ab}$ are the GT operator single-nucleon reduced matrix elements, and $R^{p}_{a}(r)$ and $R^{n}_{b}(r)$ single-proton and neutron radial harmonic oscillator wavefunctions. Previous NSM weak-decay studies do not suggest qualitative differences when using more realistic radial bases, such as Woods--Saxon~\cite{Wang:2019hjy,Weiss:2021rig}.

We perform nine different NSM calculations for the radial GT transition densities.  First, we use a valence space, consisting of the $1p_{3/2}$, $0f_{5/2}$, $1p_{1/2}$ and $0g_{9/2}$ orbitals (in $nl_j$ notation) for both protons and neutrons, with four different isospin-symmetric NSM Hamiltonians: JUN45~\cite{Honma2009}, GCN2850~\cite{Caurier2008}, RG.PROLATE~\cite{PhysRevC.80.048501} and JJ4BB~\cite{Lisetskiy2004}. While all these Hamiltonians describe well the structure of nuclei in the mass region of $^{71}$Ga and $^{71}$Ge, their predictions for weak decays can differ substantially~\cite{Castillo:2025wyr,Horoi:2026fxp}. In addition, for each Hamiltonian, we use both the bare GT operator and the effective GT operator from Ref.~\cite{Coraggio:2018tuo}, evolved consistently for this valence space. Furthermore, to explore sensitivity to the valence space, we extend it to the $0f_{7/2}$ orbital for both protons and neutrons, using the LNPS interaction~\cite{Lenzi:2010zzb}. In order to make the latter calculation feasible, we limit the occupation of the $0g_{9/2}$ orbital to two neutron excitations. With this, we investigate all valence spaces considered in previous NSM studies of the gallium anomaly~\cite{Kostensalo:2019vmv,Haxton:2025hye,Iwata:2025tzk}. 
We perform all calculations with the NSM code NATHAN~\cite{Caurier2005,Caurier:1999}.

\begin{figure}[t]
    \centering
    \includegraphics[width=\columnwidth]{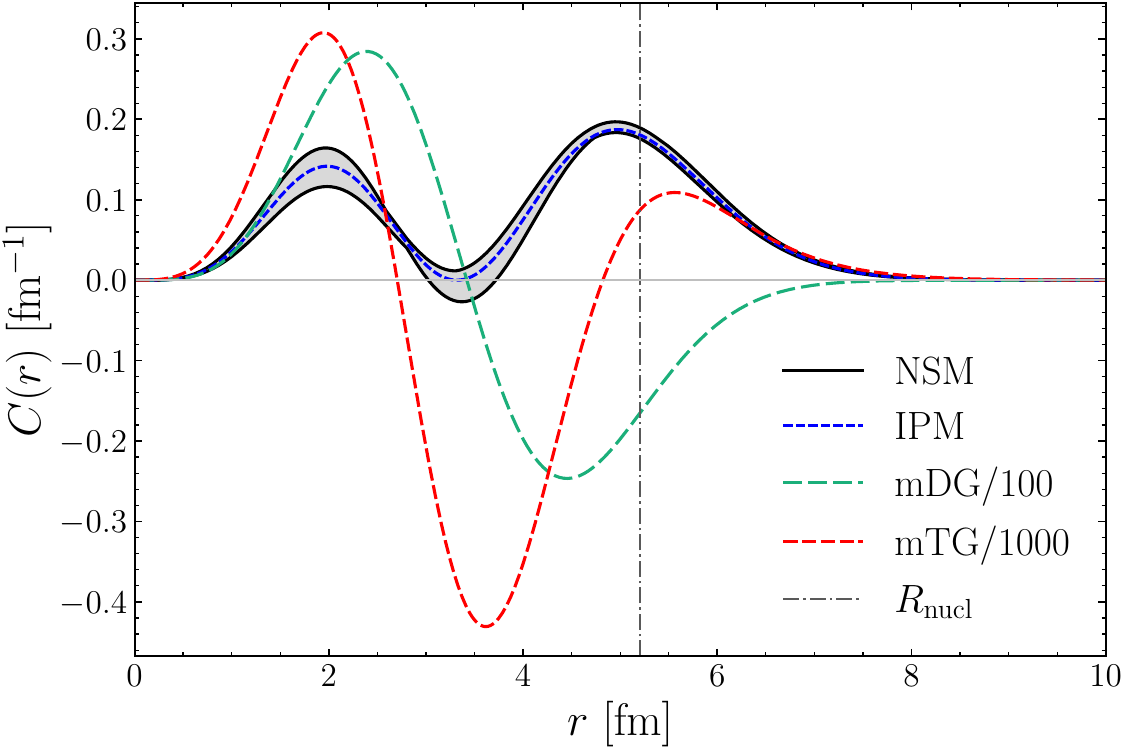}
    \caption{GT radial nuclear transition densities $C(r)$ for $^{71}\mathrm{Ga}\leftrightarrow{}^{71}\mathrm{Ge}$ EC and ${\rm I\beta D}$. The gray band spans the densities obtained in nine NSM calculations (see text), and the dashed blue line indicates the  independent particle model (IPM) density. The green and red dashed lines represent the mDG and mTG transition densities from Ref.~\cite{Cadeddu:2025ueh}, rescaled by factors 100 and 1000, respectively. The vertical line marks $R_{\rm nucl}=\sqrt{5/3}\,R_{\rm ch}$.}
    \label{fig:densities}
\end{figure}

\section{Results and discussion}
Figure~\ref{fig:densities} shows the calculated GT radial transition densities as a function of the distance $r$ from the centre of the nucleus. 
The gray band spans the NSM results for all nine calculations described above. The narrow width of the band highlights that the NSM transition density shape is robust with respect to the choice of valence space, nuclear Hamiltonian, and GT operator. The dashed blue line corresponds to the dominant $ 1p_{1/2}\leftrightarrow 1p_{3/2}$ transition, the sole contribution in the independent-particle model.

In addition, Fig.~\ref{fig:densities} compares nuclear transition densities from Ref.~\cite{Cadeddu:2025ueh}, which solve the gallium anomaly without extending significantly beyond the nuclear surface, roughly indicated by the vertical line. These densities are labeled as modified double Gaussian (mDG, green dashed curve) and modified triple Gaussian (mTG, red dashed curve). They are not obtained from a many-body calculation; instead, these densities are parameterized by analytic functions and fitted to reproduce the measured $^{71}$Ge half-life and $^{71}$Ga I$\beta$D cross sections, resolving the gallium anomaly by construction. Figure~\ref{fig:densities} highlights that the main difference between the NSM and both fitted densities---mDG and mTG---is that the latter have overall scales 100 and 1000 times larger, respectively, making them very different from the calculated nuclear transition densities. Such large discrepancy is very similar for the NSM densities before multiplying by $q_\text{EC}$ to reproduce the EC half-life. Figure~\ref{fig:densities} also shows that the NSM densities do not develop a sizeable node, in contrast to the mDG and mTG cases.  


\begin{figure}[t]
    \centering
    \includegraphics[width=\columnwidth]{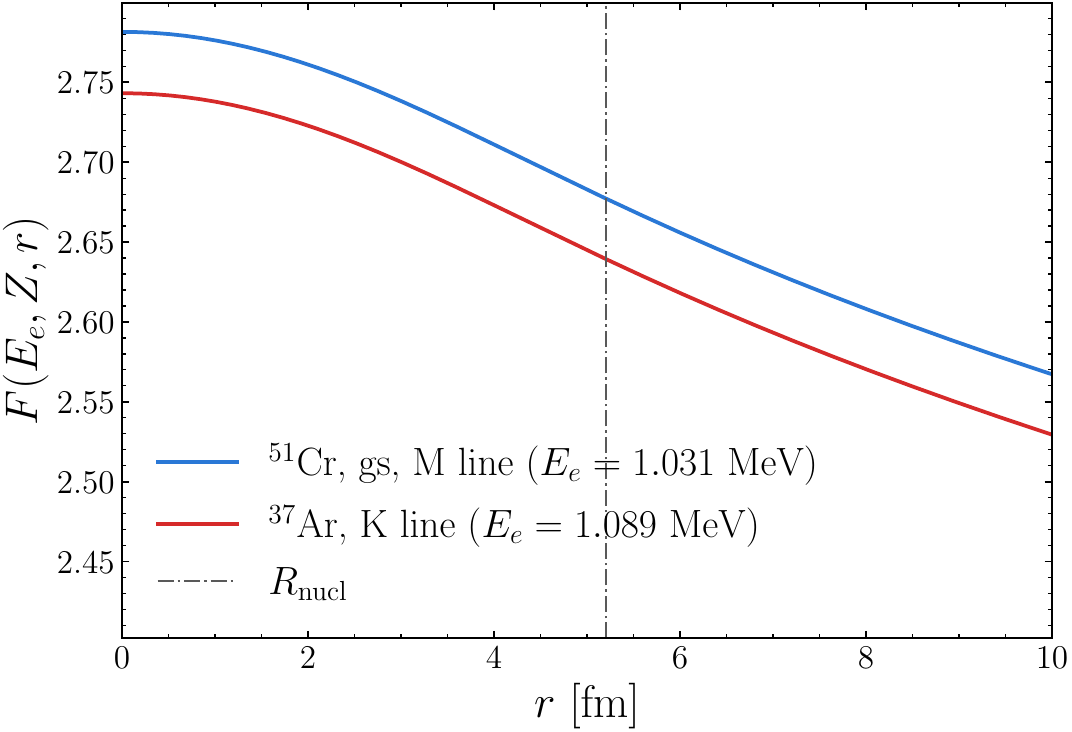}
    \caption{Fermi function $F(E_e,r)=g_{-1}^2(r)+f_{1}^2(r)$ as a function of the distance $r$ from the center of the nucleus. 
    The blue (red) line corresponds to the outgoing electron, with energy $E_e$, produced from $M$-shell ($K$-shell) capture on $^{71}$Ga of $^{51}$Cr ($^{37}$Ar) ground-state (gs) $\nu_e$'s. The vertical line indicates the nuclear radius $R_{\rm nucl}$. }
    \label{fig:Fermi_function}
\end{figure}

\begin{table*}[t]
    \centering
    \renewcommand{\arraystretch}{1.4}  
    \setlength{\tabcolsep}{8pt}        
    \begin{tabular}{ccccccc}
    \hline \hline
       \multirow{1}{*}{} & \multicolumn{2}{c}{$\sigma(^{51}\mathrm{Cr})$ [$10^{-45}$ cm$^2$]} & \multicolumn{2}{c}{$\sigma(^{37}\mathrm{Ar})$ [$10^{-45}$ cm$^2$]} & \multicolumn{2}{c}{$t_{1/2}$ [d]} \\
       \hline
       Exp. & \multicolumn{2}{c}{$4.44\pm0.33$}\,\cite{GALLEX:1992gcp, hampel_gallex_1999, Kaether:2010ag, GNO:2005bds, SAGE:2009eeu, doi:10.1142/9789811204296_0002, Barinov:2021asz, Barinov:2022wfh, cadeddu_reassessing_2026} & \multicolumn{2}{c}{$5.21\pm0.72$}\,\cite{SAGE:2009eeu, doi:10.1142/9789811204296_0002, cadeddu_reassessing_2026} & \multicolumn{2}{c}{$11.465\pm0.003$}\,\cite{cadeddu_reassessing_2026, collar_new_2023, Hampel1985, Norman2024, Derbin2025} \\
        \hline
       Scheme & BF & WF & BF & WF & BF & WF \\
       \hline
        NSM & $5.26\pm0.01$ & 5.21 & $6.30\pm0.01$ & 6.24 & $11.462\pm0.011$ & 11.465 \\
        mDG~\cite{Cadeddu:2025ueh} & 4.50 & 17.15 & 5.24 & 20.53 & 11.465 & 3.505 \\
        mTG~\cite{Cadeddu:2025ueh} & 4.39 & 35.63 & 5.10 & 42.65 & 11.463 & 1.687 \\
    \hline \hline
    \end{tabular}
    \caption{ Ground-state $^{71}$Ga I$\beta$D cross sections for $^{51}$Cr and $^{37}$Ar $\nu_e$ sources, and $^{71}$Ge EC half-life. Theoretical beyond (BF) and within (WF) factorization scheme results are compared with experimental weighted-average~\cite{cadeddu_reassessing_2026} half-lives~\cite{collar_new_2023, Hampel1985, Norman2024, Derbin2025} and cross sections from GALLEX~\cite{GALLEX:1992gcp, hampel_gallex_1999, Kaether:2010ag, GNO:2005bds}, SAGE~\cite{SAGE:2009eeu, doi:10.1142/9789811204296_0002}, and BEST~\cite{Barinov:2021asz, Barinov:2022wfh}. The NSM values are the average and spread of nine calculations (see text), and are 
    fixed to reproduce the experimental half-life in the factorization scheme. The mDG and mTG rows use the best-fit transition densities of Ref.~\cite{Cadeddu:2025ueh}.}
    \label{tab:results}
\end{table*}


Figure~\ref{fig:Fermi_function} shows the lepton part of the $\nu$ capture cross section, the continuum electron wavefunction given by the Fermi function $F(E_e,r)=g_{-1}^2(r)+f_{1}^2(r)$---which encompasses the dominant contributions in Eq.~\eqref{eq:IBD}---for two atomic shells: $M$-shell capture on $^{71}$Ga from $^{51}$Cr $\nu_e$'s (blue curve) and $K$-shell capture from $^{37}$Ar $\nu_e$'s (red curve). In both cases, $F(r)$ varies only weakly inside the nucleus, decreasing by about $3\%$ between $r=0$ and $r=R_{\rm nucl}$. The electron wavefunctions are therefore nearly constant over the region where $C(r)$ is nonzero. This is the main motivation for the factorization of the lepton and nuclear wavefunctions.

We then combine the results of Figs.~\ref{fig:densities} and~\ref{fig:Fermi_function} and calculate the ground-state $^{71}$Ga I$\beta$D cross sections for $^{51}$Cr and $^{37}$Ar sources, and the $^{71}$Ge half-life. We obtain results both beyond and within factorization, testing the validity of this scheme. 
Table~\ref{tab:results} compares the theoretical results for the different nuclear transition densities in Fig.~\ref{fig:densities} with the experimental values. The mDG and mTG results are extracted from Ref.~\cite{Cadeddu:2025ueh}. The uncertainty in the beyond-factorization NSM results reflects the spread among the nine NSM values. Within the factorization scheme, however, every NSM calculation reproduces the measured half-life once $q_{\rm EC}$ is included, and also the cross sections are common as they only depend on $M_\text{GT}$. 

Table~\ref{tab:results} indicates that the NSM cross sections are systematically larger than the experimental values, so the gallium anomaly remains unresolved. Our results also show that, when using NSM transition densities, the factorized and beyond-factorization cross sections agree at the percent level, supporting the detailed balance assumption. This is consistent with the recent analysis of Ref.~\cite{Elliott:2026zsc}, which shows that for solar neutrinos as a source675aviemren
the measured gallium capture rate agrees with the predicted one based on the factorized detailed-balance cross section. Reference~\cite{Cadeddu:2026apw} also supports the factorization scheme for $^{71}$Ge EC ratios. 

In contrast, the mDG and mTG results in Table~\ref{tab:results} change dramatically between the two schemes, with cross sections (half-lives) larger (shorter) by factors of about 4 and 8, respectively, when factorization is imposed. Such a sizeable deviation from the factorized result, given the rather constant $F(r)$ illustrated in Fig.~\ref{fig:Fermi_function}, stems from the very large variation of $C(r)$ across the nuclear surface, a factor 100 or 1000 larger than in NSM transition densities. Therefore, the breakdown of the factorization behind the detailed-balance assumption requires anomalously large-scale nuclear transition densities. At the same time, the order of magnitude of the EC half-life and the I$\beta$D cross-section are set by the $M_\text{GT}$ nuclear matrix element. Accordingly, the integral over $r$ of all transition densities in Fig.~\ref{fig:densities} is of the same order. This explains the sizeable nodes in the fitted densities from Ref.~\cite{Cadeddu:2025ueh}: they lead to a fine-tuned cancellation between the positive and negative lobes, so that the radial integral of the transition density only amounts to about $2.5\%$ (mDG) and $0.25\%$ (mTG) of the integral of its absolute value. In contrast, there is no cancellation for the NSM densities.



\begin{figure}[b]
    \centering
    \includegraphics[width=\columnwidth]{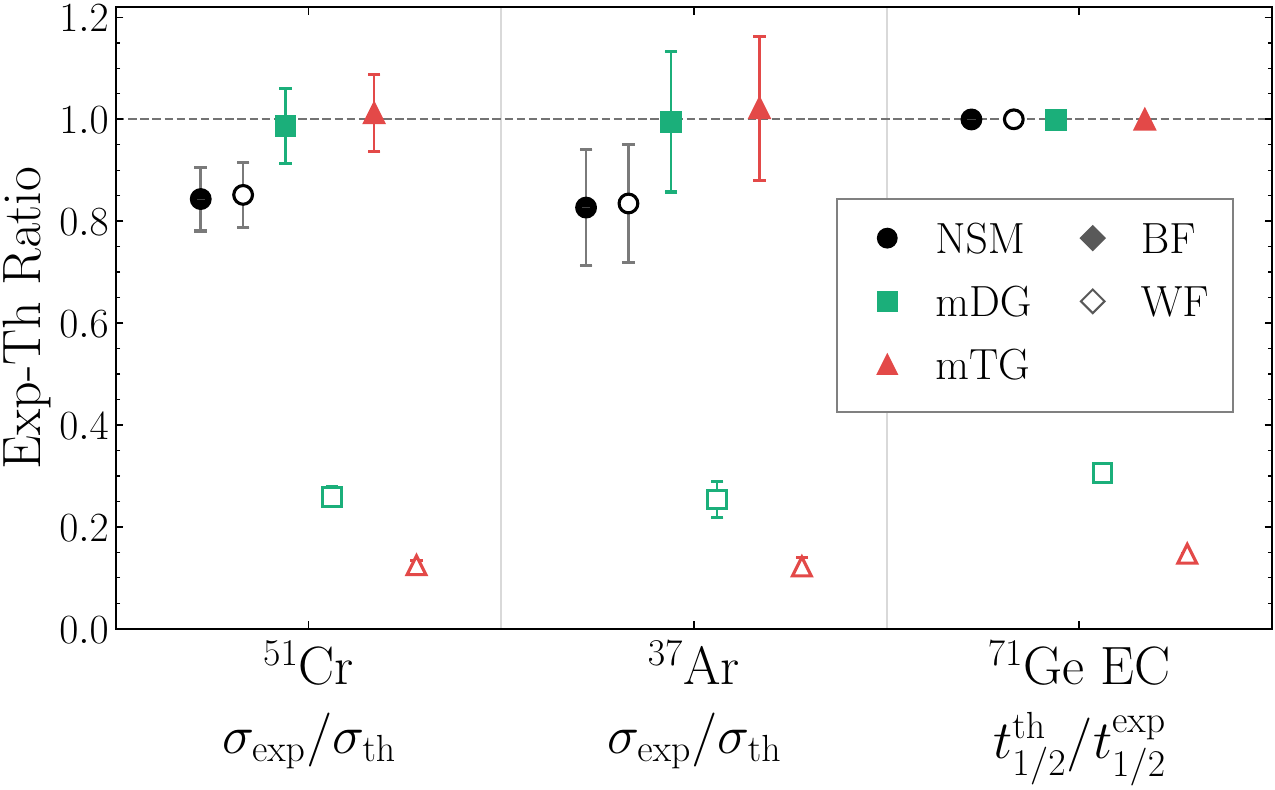}
    \caption{Experiment-to-theory ratio of the $^{71}$Ga I$\beta$D cross section for $^{51}$Cr and $^{37}$Ar sources,
    and the $^{71}$Ge EC half-life. Ratios for NSM (black), mDG (green)~\cite{Cadeddu:2025ueh} and mTG (red)~\cite{Cadeddu:2025ueh} results obtained beyond (BF, filled symbols) and within (WF, open symbols) the factorization scheme.
    }
    \label{fig:Ratio}
\end{figure}

Figure~\ref{fig:Ratio} summarizes the results of Table~\ref{tab:results}, comparing the experiment-to-theory ratios for the $^{71}$Ga $\nu_e$ capture cross sections $\sigma(^{51}\mathrm{Cr})$, $\sigma(^{37}\mathrm{Ar})$ and the $^{71}$Ge EC half-life. Ratios are given for the NSM (black), mDG (green) and mTG (red) transition densities, using the beyond (filled symbols) and within (empty ones) factorization schemes. The horizontal line indicates perfect agreement between experiment and theory. The experimental uncertainty dominates the error bars, except for the NSM half-life beyond factorization, where the spread among NSM results is larger.  By construction, the NSM and beyond-factorization fitted densities reproduce the half-life. However, for the cross sections, both theoretical NSM ratios are about $0.83$, exceeding the measured values by about $20\%$, regardless of whether factorization is assumed. In contrast, the ratios obtained with the fitted densities of Ref.~\cite{Cadeddu:2025ueh} drop to about $0.12$--$0.26$ with factorization, but beyond this scheme they agree with experiment.

Finally, we test whether a NSM-like transition density can solve the gallium anomaly. We repeat the fit of Ref.~\cite{Cadeddu:2025ueh} with a NSM radial ansatz instead of ad hoc mDG and mTG functions. Since the bare GT operator only connects single-nucleon orbitals with the same $n$ and $l$---additional contributions allowed by the effective GT operator are minor---the NSM transition density is a linear combination of radial functions of active orbitals:
\begin{equation}
C(r)=\sum_{nl} c_{nl}\,R_{nl}^2(r)\,r^2\, .
\label{eq:fitted}
\end{equation}
The coefficients $c_{nl}$ collect the OBTDs and angular factors for all orbitals with the same $nl$, which in the valence spaces considered here are $0f$, $1p$, and $0g$. We treat the coefficients as free parameters, and find that reproducing simultaneously the measured EC half-life and $^{51}$Cr and $^{37}$Ar cross sections requires fitted $c_{nl}$ values roughly 3000 times larger than the calculated NSM ones. Further, this fitted density is fine tuned so that its positive and negative lobes cancel at the percent level. Both features are consistent with the densities from Ref.~\cite{Cadeddu:2025ueh}. In conclusion, fitted transition densities from Eq.~\eqref{eq:fitted} that solve the gallium anomaly cannot be realized within the NSM.

\section{ Summary}
In this work we calculate the $^{71}$Ga $\nu_e$ capture cross section and the $^{71}$Ge EC half-life in the beyond-factorization scheme proposed in Ref.~\cite{Cadeddu:2025ueh}. We obtain continuum (for $\nu_e$ capture cross section) and bound (for EC) electron wavefunctions within the Dirac–Hartree–Fock–Slater approach. We also calculate nine NSM transition densities, using five nuclear Hamiltonians, bare and effective GT operators, and exploring the $pf$ and $0g_{9/2}$ orbital configuration spaces. After normalizing each calculation to exactly reproduce the $^{71}$Ge EC half-life, the resulting nuclear GT transition densities are remarkably robust against the choice of Hamiltonian, operator, and valence space.

With our NSM transition densities, the $^{71}$Ga neutrino-capture cross section remains about $20\%$ larger than the measured value, which does not resolve the gallium anomaly. In contrast to the results of Ref.~\cite{Cadeddu:2025ueh}, obtained with ad hoc fitted nuclear transition densities, our beyond- and within-factorization results agree at the $1\%$ level. This finding is consistent with the solar-neutrino analysis of Ref.~\cite{Elliott:2026zsc}, which supports the detailed-balance cross section in these two weak processes. We also explore whether transition densities consistent with NSM single-nucleon radial densities can explain the anomaly. We find that a solution requires fine-tuned densities with an overall scale several orders of magnitude larger than realistic NSM ones, and a percent-level cancellation of the GT matrix element absent in NSM results. These features, disfavored by our calculations, are also common to the fitted nuclear transition densities of Ref.~\cite{Cadeddu:2025ueh} that solve the gallium anomaly.


Future work can extend the current analysis to ab initio approaches, similar to the NSM-ab initio comparisons of light-nuclei weak decays that have indicated mildly different theoretical transition densities~\cite{Wang:2019hjy,Weiss:2021rig}. In the heavier Ga and Ge systems, the valence-space in-medium similarity renormalization group~\cite{Stroberg:2019mxo} would allow the use of consistent one- and two-body currents as well as effective operators~\cite{Gysbers2019,Stroberg2021,Li:2025exk}. In addition, calculations with other many-body methods~\cite{Ney:2021anu, Horiuchi:2021pup} would provide a valuable test of the conclusions presented in this work.

\section*{Acknowledgements}
We thank José María Fernández Varea for useful discussions. This work is financially supported by MCIN/AEI/10.13039/501100011033 from the following grants: PID2023-147112NB-C22, PREP2023-002179 Doctoral Grant cofunded by FSE+ 2021–2027 and CEX2024-001451-M to the “Unit of Excellence María de Maeztu 2025–2031” award to the Institute of Cosmos Sciences; and by the Generalitat de Catalunya, through grant 2021SGR01095 and the predoctoral program AGAUR-FI ajuts Joan Oró 2025 FI (2025 FI-1 01213) de la Secretaria d’Universitats i Recerca cofunded by FSE+ 2021–2027.

\bibliographystyle{apsrev4-2}
\bibliography{reference}

@article{Elliott:2023cvh,
    author = "Elliott, Steven R. and Gavrin, Vladimir and Haxton, Wick",
    title = "{The gallium anomaly}",
    primaryClass = "nucl-ex",
    doi = "10.1016/j.ppnp.2023.104082",
    journal = "Prog. Part. Nucl. Phys.",
    volume = "134",
    pages = "104082",
    year = "2024"
}

@article{Giunti:2010zu,
    author = "Giunti, Carlo and Laveder, Marco",
    title = "{Statistical Significance of the Gallium Anomaly}",
    primaryClass = "hep-ph",
    doi = "10.1103/PhysRevC.83.065504",
    journal = "Phys. Rev. C",
    volume = "83",
    pages = "065504",
    year = "2011"
}

@article{GALLEX:1992gcp,
    author = "Anselmann, P. and others",
    collaboration = "GALLEX",
    title = "{Solar neutrinos observed by GALLEX at Gran Sasso.}",
    doi = "10.1016/0370-2693(92)91521-A",
    journal = "Phys. Lett. B",
    volume = "285",
    pages = "376--389",
    year = "1992"
}

@article{Kaether:2010ag,
    author = "Kaether, F. and Hampel, W. and Heusser, G. and Kiko, J. and Kirsten, T.",
    title = "{Reanalysis of the GALLEX solar neutrino flux and source experiments}",
    primaryClass = "hep-ex",
    doi = "10.1016/j.physletb.2010.01.030",
    journal = "Phys. Lett. B",
    volume = "685",
    pages = "47--54",
    year = "2010"
}

@article{GNO:2005bds,
    author = "Altmann, M. and others",
    collaboration = "GNO",
    title = "{Complete results for five years of GNO solar neutrino observations}",
    reportNumber = "MPI-K-2005-008",
    doi = "10.1016/j.physletb.2005.04.068",
    journal = "Phys. Lett. B",
    volume = "616",
    pages = "174--190",
    year = "2005"
}

@article{SAGE:2009eeu,
    author = "Abdurashitov, J. N. and others",
    collaboration = "SAGE",
    title = "{Measurement of the solar neutrino capture rate with gallium metal. III: Results for the 2002--2007 data-taking period}",
    primaryClass = "nucl-ex",
    doi = "10.1103/PhysRevC.80.015807",
    journal = "Phys. Rev. C",
    volume = "80",
    pages = "015807",
    year = "2009"
}

@inbook{doi:10.1142/9789811204296_0002,
author = {V. N. Gavrin},
title = {The history, present and future of SAGE (Soviet-American Gallium Experiment)},
booktitle = {Solar Neutrinos},
chapter = {},
pages = {29-46},
year={2019},
editorial = {M. Meyer & K. Zuber}, 
doi = {10.1142/9789811204296_0002},
URL = {https://www.worldscientific.com/doi/abs/10.1142/9789811204296_0002},
}

@article{hampel_gallex_1999,
  title = {{GALLEX} solar neutrino observations: results for {GALLEX} {IV}},
  volume = {447},
  issn = {0370-2693},
  url = {https://www.sciencedirect.com/science/article/pii/S0370269398015792},
  doi = {https://doi.org/10.1016/S0370-2693(98)01579-2},
  number = {1},
  journal = {Phys. Lett. B},
  author = {Hampel, W. and Handt, J. and Heusser, G. and others},
  year = {1999},
  pages = {127--133},
}

@article{Barinov:2021asz,
    author = "Barinov, V. V. and others",
    title = "{Results from the Baksan Experiment on Sterile Transitions (BEST)}",
    primaryClass = "nucl-ex",
    doi = "10.1103/PhysRevLett.128.232501",
    journal = "Phys. Rev. Lett.",
    volume = "128",
    number = "23",
    pages = "232501",
    year = "2022"
}

@article{Barinov:2022wfh,
    author = "Barinov, V. V. and others",
    title = "{Search for electron-neutrino transitions to sterile states in the BEST experiment}",
    primaryClass = "nucl-ex",
    doi = "10.1103/PhysRevC.105.065502",
    journal = "Phys. Rev. C",
    volume = "105",
    number = "6",
    pages = "065502",
    year = "2022"
}

@article{Elliott:2023xkb,
    author = "Elliott, S. R. and Gavrin, V. N. and Haxton, W. C. and Ibragimova, T. V. and Rule, E. J.",
    title = "{Gallium neutrino absorption cross section and its uncertainty}",
    primaryClass = "nucl-th",
    reportNumber = "N3AS-23-009",
    doi = "10.1103/PhysRevC.108.035502",
    journal = "Phys. Rev. C",
    volume = "108",
    number = "3",
    pages = "035502",
    year = "2023"
}

@article{Haxton:2025hye,
    author = "Haxton, W. C. and Rule, Evan",
    title = "{The gallium solar neutrino capture cross section revisited}",
    primaryClass = "nucl-ex",
    reportNumber = "LA-UR-24-32851, N3AS-24-043",
    doi = "10.1016/j.physletb.2025.139259",
    journal = "Phys. Lett. B",
    volume = "861",
    pages = "139259",
    year = "2025"
}

@article{Kostensalo:2019vmv,
    author = "Kostensalo, Joel and Suhonen, Jouni and Giunti, Carlo and Srivastava, Praveen C.",
    title = "{The gallium anomaly revisited}",
    primaryClass = "nucl-th",
    doi = "10.1016/j.physletb.2019.06.057",
    journal = "Phys. Lett. B",
    volume = "795",
    pages = "542--547",
    year = "2019",
    note = "[Erratum: Phys.Lett.B 846, 138190 (2023)]"
}

@article{Iwata:2025tzk,
    author = "Iwata, Yoritaka and Ejiri, Hiroyasu and Sarkar, Shahariar",
    title = "{Low-energy neutrino responses for Ga71 by electron capture rates, charge exchange reactions, and nuclear shell model calculations}",
    primaryClass = "nucl-th",
    doi = "10.1103/7cw1-f5dj",
    journal = "Phys. Rev. C",
    volume = "114",
    number = "2",
    pages = "024317",
    year = "2026"
}

@article{GiuntiLaveder2007,
  author  = {Giunti, C. and Laveder, M.},
  title   = {Short-Baseline Active-Sterile Neutrino Oscillations?},
  journal = {Mod. Phys. Lett. A},
  volume  = {22},
  pages   = {2499--2509},
  year    = {2007},
}

@article{GiuntiLasserre2019,
  author  = {Giunti, C. and Lasserre, T.},
  title   = {eV-Scale Sterile Neutrinos},
  journal = {Ann. Rev. Nucl. Part. Sci.},
  volume  = {69},
  pages   = {163--190},
  year    = {2019},
  doi     = {10.1146/annurev-nucl-101918-023755},
}

@article{Gariazzo2016,
  author  = {Gariazzo, S. and Giunti, C. and Laveder, M.
             and Li, Y. F. and Zavanin, E. M.},
  title   = {Light Sterile Neutrinos},
  journal = {J. Phys. G},
  volume  = {43},
  pages   = {033001},
  year    = {2016},
  doi     = {10.1088/0954-3899/43/3/033001},
}

@article{GonzalezGarcia2016,
  author  = {Gonzalez-Garcia, M. C. and Maltoni, M.
             and Schwetz, T.},
  title   = {Global Analyses of Neutrino Oscillation Experiments},
  journal = {Nucl. Phys. B},
  volume  = {908},
  pages   = {199--217},
  year    = {2016},
  doi     = {10.1016/j.nuclphysb.2016.02.033},
}

@article{Diaz2020,
  author  = {Diaz, A. and Argüelles, C. A. and Collin, G. H.
             and Conrad, J. M. and Shaevitz, M. H.},
  title   = {Where Are We With Light Sterile Neutrinos?},
  journal = {Phys. Rep.},
  volume  = {884},
  pages   = {1--59},
  year    = {2020},
  doi     = {10.1016/j.physrep.2020.08.005},
}

@article{Boser2020,
  author  = {Boser, S. and Buck, C. and Giunti, C.
             and Lesgourgues, J. and Ludhova, L.
             and Mertens, S. and Schukraft, A. and Wurm, M.},
  title   = {Status of Light Sterile Neutrino Searches},
  journal = {Prog. Part. Nucl. Phys.},
  volume  = {111},
  pages   = {103736},
  year    = {2020},
  doi     = {10.1016/j.ppnp.2019.103736},
}

@article{DasguptaKopp2021,
  author  = {Dasgupta, B. and Kopp, J.},
  title   = {Sterile Neutrinos},
  journal = {Phys. Rep.},
  volume  = {928},
  pages   = {1--63},
  year    = {2021},
  doi     = {10.1016/j.physrep.2021.06.002},
}

@article{GiuntiTernes2024,
  author  = {Giunti, C. and Ternes, C. A.},
  title   = {Confronting Solutions of the Gallium Anomaly
             with Reactor Rate Data},
  journal = {Phys. Lett. B},
  volume  = {849},
  pages   = {138436},
  year    = {2024},
  doi     = {10.1016/j.physletb.2023.138436}
}

@misc{Abazajian2012,
  author        = {Abazajian, K. N. and others},
  title         = {Light Sterile Neutrinos: A White Paper},
  year          = {2012},
  eprint        = {1204.5379},
  archivePrefix = {arXiv},
  primaryClass  = {hep-ph},
  doi           = {10.48550/arXiv.1204.5379}
}

@article{Acero2024,
  author  = {Acero, M. A. and others},
  title   = {White Paper on Light Sterile Neutrino Searches
             and Related Phenomenology},
  journal = {J. Phys. G},
  volume  = {51},
  pages   = {120501},
  year    = {2024},
  doi     = {10.1088/1361-6471/ad307f},
}

@article{FarzanSchwetz2023,
  author  = {Farzan, Y. and Schwetz, T.},
  title   = {A Decoherence Explanation of the Gallium
             Neutrino Anomaly},
  journal = {SciPost Phys.},
  volume  = {15},
  pages   = {172},
  year    = {2023},
  doi     = {10.21468/SciPostPhys.15.4.172}
}

@article{Giunti2022Gallium,
  author  = {Giunti, C. and Li, Y. F. and Ternes, C. A.
             and Tyagi, O. and Xin, Z.},
  title   = {Gallium Anomaly: Critical View from the Global
             Picture of $\nu_e$ and $\bar{\nu}_e$ Disappearance},
  journal = {J. High Ener. Phys.},
  volume  = {2022},
  number  = {10},
  pages   = {164},
  year    = {2022},
  doi     = {10.1007/JHEP10(2022)164},
  primaryClass  = {hep-ph}
}

@article{Giunti2022Reactor,
  author  = {Giunti, C. and Li, Y. F. and Ternes, C. A.
             and Xin, Z.},
  title   = {Reactor Antineutrino Anomaly in Light of Recent
             Flux Model Refinements},
  journal = {Phys. Lett. B},
  volume  = {829},
  pages   = {137054},
  year    = {2022},
  doi     = {10.1016/j.physletb.2022.137054},
  primaryClass  = {hep-ph}
}

@article{STEREO2023,
  author  = {{The STEREO Collaboration}},
  title   = {STEREO Neutrino Spectrum of $^{235}$U Fission
             Rejects Sterile Neutrino Hypothesis},
  journal = {Nature},
  volume  = {613},
  pages   = {257--261},
  year    = {2023},
  doi     = {10.1038/s41586-022-05568-2},
  primaryClass  = {hep-ex}
}

@article{Machikhiliyan2022,
  author  = {Machikhiliyan, I.},
  title   = {The DANSS Neutrino Spectrometer: The Results
             of Reactor Antineutrino Studies},
  journal = {Phys. Part. Nucl.},
  volume  = {53},
  pages   = {546--551},
  year    = {2022},
  doi     = {10.1134/S1063779622020502}
}

@article{PROSPECT2025,
  author  = {Andriamirado, M. and others},
  collaboration = {PROSPECT},
  title   = {Final Search for Short-Baseline Neutrino
             Oscillations with the PROSPECT-I Detector at HFIR},
  journal = {Phys. Rev. Lett.},
  volume  = {134},
  number  = {15},
  pages   = {151802},
  year    = {2025},
  doi     = {10.1103/PhysRevLett.134.151802},
  primaryClass  = {hep-ex}
}

@article{Goldhagen2022,
  author  = {Goldhagen, K. and Maltoni, M.
             and Reichard, S. E. and Schwetz, T.},
  title   = {Testing sterile neutrino mixing with present
             and future solar neutrino data},
  journal = {Eur. Phys. J. C},
  volume  = {82},
  pages   = {116},
  year    = {2022},
  doi     = {10.1140/epjc/s10052-022-10052-2}
}

@article{GonzalezGarcia2025,
  author  = {Gonzalez-Garcia, M. C. and Maltoni, M.
             and Pinheiro, J. P.},
  title   = {Solar model independent constraints on the
             sterile neutrino interpretation of the
             Gallium Anomaly},
  journal = {Phys. Lett. B},
  volume  = {862},
  pages   = {139297},
  year    = {2025},
  doi     = {10.1016/j.physletb.2025.139297}
}

@article{MicroBooNE2025,
  author  = {Abratenko, P. and others},
  collaboration = {MicroBooNE},
  title   = {Search for light sterile neutrinos with two
             neutrino beams at MicroBooNE},
  journal = {Nature},
  volume  = {648},
  pages   = {64--69},
  year    = {2025},
  doi     = {10.1038/s41586-025-09757-7}
}

@article{KATRIN2025,
  author  = {Acharya, H. and others},
  collaboration = {KATRIN},
  title   = {Sterile-neutrino search based on 259 days
             of KATRIN data},
  journal = {Nature},
  volume  = {648},
  pages   = {70--75},
  year    = {2025},
  doi     = {10.1038/s41586-025-09739-9}
}

@article{Huber2025,
  author  = {Huber, P.},
  title   = {Still no sign of hypothetical sterile-neutrino particle},
  journal = {Nature},
  volume  = {648},
  pages   = {40--42},
  year    = {2025},
  doi     = {10.1038/d41586-025-03726-w}
}

@article{Bahcall1997,
  author        = {Bahcall, J. N.},
  title         = {Gallium Solar Neutrino Experiments:
                   Absorption Cross Sections, Neutrino
                   Spectra, and Predicted Event Rates},
  journal       = {Phys. Rev. C},
  volume        = {56},
  pages         = {3391--3409},
  year          = {1997},
  doi           = {10.1103/PhysRevC.56.3391},
 
}

@book{Bahcall1989,
  author    = {Bahcall, J. N.},
  title     = {Neutrino Astrophysics},
  publisher = {Cambridge University Press},
  address   = {Cambridge, New York},
  year      = {1989},
  isbn      = {9780521351133}
}

@article{Bahcall1978,
  author  = {Bahcall, J. N.},
  title   = {Solar Neutrino Experiments},
  journal = {Rev. Mod. Phys.},
  volume  = {50},
  pages   = {881--908},
  year    = {1978},
  doi     = {10.1103/RevModPhys.50.881}
}

@article{Bahcall1964,
  author  = {Bahcall, J. N.},
  title   = {Solar Neutrino Cross Sections and Nuclear Beta Decay},
  journal = {Phys. Rev.},
  volume  = {135},
  pages   = {B137--B146},
  year    = {1964},
  doi     = {10.1103/PhysRev.135.B137}
}

@article{Cadeddu:2025ueh,
    author = "Cadeddu, M. and Cargioli, N. and Dordei, F. and Ferro, L. and Giunti, C. and Pitzalis, M.",
    title = "{Possible Solution to the Gallium Anomaly Moving beyond the Leptonic Wave-Function Factorization}",
    primaryClass = "hep-ph",
    doi = "10.1103/5pvf-mcr1",
    journal = "Phys. Rev. Lett.",
    volume = "137",
    number = "13",
    pages = "131805",
    year = "2026"
}

@misc{Elliott:2026zsc,
    author = "Elliott, Steven R. and Robertson, R. G. Hamish",
    title = "{Gallium Anomaly and Solar Neutrinos}",
    eprint = "2610.22219",
    archivePrefix = "arXiv",
    primaryClass = "nucl-ex",
    month = "10",
    year = "2026"
}

@article{salvat_radial_2019,
  title = {radial: {A} {Fortran} subroutine package for the solution of the radial {Schrödinger} and {Dirac} wave equations},
  volume = {240},
  issn = {0010-4655},
  url = {https://www.sciencedirect.com/science/article/pii/S0010465519300633},
  doi = {https://doi.org/10.1016/j.cpc.2019.02.011},
  
  journal = {Comp. Phys. Comm.},
  author = {Salvat, Francesc and Fernández-Varea, José M.},
  year = {2019},
  pages = {165--177},
}

@article{cadeddu_reassessing_2026,
  title = {Reassessing the gallium anomaly using self-consistent electron wave functions},
  volume = {113},
  issn = {2470-0010, 2470-0029},
  url = {https://link.aps.org/doi/10.1103/pz8g-zz1b},
  doi = {10.1103/pz8g-zz1b},
  
  language = {en},
  number = {3},
  urldate = {2026-10-13},
  journal = {Phys. Rev. D},
  author = {Cadeddu, M. and Cargioli, N. and Carotenuto, G. and Dordei, F. and Ferro, L. and Giunti, C.},
  month = feb,
  year = {2026},
  pages = {033006},
}

@misc{Cadeddu:2026apw,
    author = "Cadeddu, M. and Cargioli, N. and Cau, M. and Dordei, F. and Ferro, L.",
    eprint = "2608.00120",
    archivePrefix = "arXiv",
    primaryClass = "hep-ph",
    month = "7",
    year = "2026"
}

@article{navas_review_2024,
  title = {Review of {Particle} {Physics}},
  volume = {110},
  url = {https://link.aps.org/doi/10.1103/PhysRevD.110.030001},
  doi = {10.1103/PhysRevD.110.030001},
  number = {3},
  journal = {Phys. Rev. D},
  publisher = {American Physical Society},
  author = {Navas, S. and Amsler, C. and Gutsche, T. and others},
  month = aug,
  year = {2024},
  pages = {030001},
}

@article{markisch_measurement_2019,
  title = {Measurement of the {Weak} {Axial}-{Vector} {Coupling} {Constant} in the {Decay} of {Free} {Neutrons} {Using} a {Pulsed} {Cold} {Neutron} {Beam}},
  volume = {122},
  issn = {0031-9007, 1079-7114},
  url = {https://link.aps.org/doi/10.1103/PhysRevLett.122.242501},
  doi = {10.1103/PhysRevLett.122.242501},
  language = {en},
  number = {24},
  urldate = {2026-10-26},
  journal = {Phys. Rev. Lett.},
  author = {Märkisch, B. and Mest, H. and Saul, H. and Wang, X. and Abele, H. and Dubbers, D. and Klopf, M. and Petoukhov, A. and Roick, C. and Soldner, T. and Werder, D.},
  month = jun,
  year = {2019},
  pages = {242501},
}

@article{Bambynek:1977zz,
    author = "Bambynek, W. and Behrens, H. and Chen, M. H. and Crasemann, B. and Fitzpatrick, M. L. and Ledingham, K. W. D. and Genz, H. and Mutterer, M. and Intemann, R. L.",
    title = "{Orbital electron capture by the nucleus}",
    doi = "10.1103/RevModPhys.49.77",
    journal = "Rev. Mod. Phys.",
    volume = "49",
    pages = "77--221",
    year = "1977",
    note = "[Erratum: Rev.Mod.Phys. 49, 961--962 (1977)]"
}

@article{EDELWEISS:2020fxc,
    author = "Arnaud, Q. and others",
    collaboration = "EDELWEISS",
    title = "{First germanium-based constraints on sub-MeV Dark Matter with the EDELWEISS experiment}",
    primaryClass = "astro-ph.GA",
    doi = "10.1103/PhysRevLett.125.141301",
    journal = "Phys. Rev. Lett.",
    volume = "125",
    number = "14",
    pages = "141301",
    year = "2020"
}

@article{collar_new_2023,
  title = {New measurements of {\textasciicircum}71{\textbackslash}{mathrmGe} decay: {Impact} on the gallium anomaly},
  volume = {108},
  url = {https://link.aps.org/doi/10.1103/PhysRevC.108.L021602},
  doi = {10.1103/PhysRevC.108.L021602},
  number = {2},
  journal = {Phys. Rev. C},
  publisher = {American Physical Society},
  author = {Collar, J. I. and Yoon, S. G.},
  month = aug,
  year = {2023},
  pages = {L021602},
}

@article{Hampel1985,
  author  = {Hampel, W. and Remsberg, L. P.},
  title   = {Half-life of $^{71}$Ge},
  journal = {Phys. Rev. C},
  volume  = {31},
  pages   = {666--667},
  year    = {1985},
  doi     = {10.1103/PhysRevC.31.666}
}

@article{Norman2024,
  author  = {Norman, E. B. and Drobizhev, A. and Gharibyan, N.
             and Gregorich, K. E. and Kolomensky, Y. G.
             and Sammis, B. N. and Scielzo, N. D.
             and Shusterman, J. A. and Thomas, K. J.},
  title   = {Half-life of $^{71}$Ge and the gallium anomaly},
  journal = {Phys. Rev. C},
  volume  = {109},
  pages   = {055501},
  year    = {2024},
  doi     = {10.1103/PhysRevC.109.055501}
}

@article{Derbin2025,
  author  = {Derbin, A. and Drachnev, I. and Ivanov, D.
             and Muratova, V. and Trushin, M.
             and Unzhakov, E. and Vorobyev, A.},
  title   = {Half-life of $^{71}$Ge},
  journal = {Phys. At. Nucl.},
  volume  = {88},
  pages   = {52--57},
  year    = {2025}
}

@article{wang_nuclear_2024,
  title = {Nuclear charge radii of germanium isotopes around {N} = 40},
  volume = {856},
  issn = {0370-2693},
  url = {https://www.sciencedirect.com/science/article/pii/S0370269324004258},
  doi = {https://doi.org/10.1016/j.physletb.2024.138867},
  journal = {Phys. Lett. B},
  author = {Wang, S. J. and Kanellakopoulos, A. and Yang, X. F. and others},
  year = {2024},
  pages = {138867},
}

@article{Bahcall:1962zz,
    author = "Bahcall, John N.",
    title = "{Electron Capture and Nuclear Matrix Elements of Be-7}",
    doi = "10.1103/PhysRev.128.1297",
    journal = "Phys. Rev.",
    volume = "128",
    pages = "1297--1301",
    year = "1962"
}

@article{Brysk:1958zz,
    author = "Brysk, H. and Rose, M. E.",
    title = "{Theoretical Results on Orbital Capture}",
    doi = "10.1103/RevModPhys.30.1169",
    journal = "Rev. Mod. Phys.",
    volume = "30",
    pages = "1169--1177",
    year = "1958"
}

@article{Coraggio:2018tuo,
    author = "Coraggio, L. and De Angelis, L. and Fukui, T. and Gargano, A. and Itaco, N. and Nowacki, F.",
    title = "{Renormalization of the Gamow-Teller operator within the realistic shell model}",
    primaryClass = "nucl-th",
    doi = "10.1103/PhysRevC.100.014316",
    journal = "Phys. Rev. C",
    volume = "100",
    number = "1",
    pages = "014316",
    year = "2019"
}

@article{Caurier2005,
author = {Caurier, E. and Martínez-Pinedo, G. and Nowacki, F. and Poves, A. and Zuker, A. P. },
title = {}, 
journal = {Rev. Mod. Phys.}, 
volume = {77}, 
pages = {427}, 
year = {2005}, 
doi = {10.1103/RevModPhys.77.427}
}

@article{Brown1988,
author = {Brown, B. A. and Wildenthal, B.},
title = {}, 
journal = {Annu. Rev. Nucl. Part. Sci.}, 
volume = {38}, 
pages = {29-66}, 
year = {1988}, 
doi = {10.1146/annurev.ns.38.120188.000333}
}

@article{Otsuka2020,
author = {Otsuka, T. and Gade, A. and Sorlin, O. and Suzuki, T. and Utsuno, Y.},
title = {}, 
journal = {Rev. Mod. Phys.}, 
volume = {92}, 
pages = {015002}, 
year = {2020}, 
doi = {10.1103/RevModPhys.92.015002}
}

@article{Honma2009,
author = {Honma, M. and Otsuka, T. and Mizusaki, T. and Hjorth-Jensen, M.},
title = {}, 
journal = {Phys. Rev. C}, 
volume = {80}, 
pages = {064323}, 
year = {2009}, 
doi = {10.1103/PhysRevC.80.064323}
}

@article{Caurier2008,
author = {Caurier, E. and Menéndez, J. and Nowacki, F. and Poves, A.},
title = {}, 
journal = {Phys. Rev. Lett.}, 
volume = {100}, 
pages = {052503}, 
year = {2008}, 
doi = {10.1103/PhysRevLett.100.052503}
}

@article{Lisetskiy2004,
author = {Lisetskiy, A. F. and Brown, B. A. and Horoi, H. G. M.},
title = {}, 
journal = {Phys. Rev. C}, 
volume = {70}, 
pages = {044314}, 
year = {2004}, 
doi = {10.1103/PhysRevC.70.044314}
}

@article{PhysRevC.80.048501,
  title = {Occupancies of individual orbits, and the nuclear matrix element of the $^{76}\mathrm{Ge}$ neutrinoless $\ensuremath{\beta}\ensuremath{\beta}$ decay},
  author = {Men{\'e}ndez, J. and Poves, A. and Caurier, E. and Nowacki, F.},
  journal = {Phys. Rev. C},
  volume = {80},
  issue = {4},
  pages = {048501},
  numpages = {4},
  year = {2009},
  month = {Oct},
  publisher = {American Physical Society},
  doi = {10.1103/PhysRevC.80.048501},
  url = {https://link.aps.org/doi/10.1103/PhysRevC.80.048501}
}

@book{Suhonen2007,
	author    = {Suhonen, Jouni},
	title     = {From Nucleons to Nucleus: Concepts of Microscopic Nuclear Theory},
	series    = {Theoretical and Mathematical Physics},
	publisher = {Springer-Verlag},
	address   = {Berlin Heidelberg},
	year      = {2007},
	isbn      = {978-3-540-48859-0},
	doi       = {10.1007/978-3-540-48861-3}
}

@article{Caurier:1999,
  author  = {Caurier, E. and Nowacki, F.},
  title   = {Present Status of Shell Model Techniques},
  journal = {Acta Phys. Pol. B},
  volume  = {30},
  number  = {3},
  pages   = {705},
  year    = {1999},
  url     = {https://www.actaphys.uj.edu.pl/R/30/3/705}
}

@article{Gysbers2019,
  author  = {Gysbers, P. and others},
  title   = {Discrepancy between experimental and theoretical $\beta$-decay rates resolved from first principles},
  journal = {Nat. Phys.},
  volume  = {15},
  pages   = {428--431},
  year    = {2019},
  doi     = {10.1038/s41567-019-0450-7}
}

@article{Stroberg2021,
  author  = {Stroberg, S. R.},
  title   = {Nuclear Structure Theory and the Shell Model},
  journal = {Particles},
  volume  = {4},
  number  = {4},
  pages   = {521--535},
  year    = {2021},
  doi     = {10.3390/particles4040038}
}

@article{Stroberg:2019mxo,
    author = "Stroberg, S. Ragnar and Bogner, Scott K. and Hergert, Heiko and Holt, Jason D.",
    title = "{Nonempirical Interactions for the Nuclear Shell Model: An Update}",
    primaryClass = "nucl-th",
    doi = "10.1146/annurev-nucl-101917-021120",
    journal = "Ann. Rev. Nucl. Part. Sci.",
    volume = "69",
    pages = "307--362",
    year = "2019"
}

@article{Li:2025exk,
    author = "Li, Zhen and Miyagi, Takayuki and Schwenk, Achim",
    title = "{Ab Initio Calculations of {\ensuremath{\beta}}-Decay Half-Lives for N=50 Neutron-Rich Nuclei}",
    primaryClass = "nucl-th",
    doi = "10.1103/xjv9-t6sn",
    journal = "Phys. Rev. Lett.",
    volume = "136",
    number = "18",
    pages = "182501",
    year = "2026"
}

@article{Danilov2021,
    author         = {Danilov, M. and Skrobova, N.},
    title          = {New results from the DANSS experiment},
    journal        = {J. Phys.: Conf. Ser.},
    volume         = {2156},
    pages          = {012067},
    year           = {2021},
    eprint         = {2112.13413},
    archivePrefix  = {arXiv},
    primaryClass   = {hep-ex},
    doi            = {10.1088/1742-6596/2156/1/012067}
}

@article{Atif2022,
    author         = {Atif, Z. and others},
    title          = {Search for sterile neutrino oscillations using RENO and NEOS data},
    journal        = {Phys. Rev. D},
    volume         = {105},
    pages          = {L111101},
    year           = {2022},
    doi            = {10.1103/PhysRevD.105.L111101}
}

@article{Choi2020,
    author         = {Choi, J. H. and others},
    collaboration  = {RENO},
    title          = {Search for Sub-eV Sterile Neutrinos at RENO},
    journal        = {Phys. Rev. Lett.},
    volume         = {125},
    pages          = {191801},
    year           = {2020},
    doi            = {10.1103/PhysRevLett.125.191801}
}

@article{Lenzi:2010zzb,
    author = "Lenzi, S. M. and Nowacki, F. and Poves, A. and Sieja, K.",
    title = "{Island of inversion around Cr-64}",
    doi = "10.1103/PhysRevC.82.054301",
    journal = "Phys. Rev. C",
    volume = "82",
    pages = "054301",
    year = "2010"
}

@article{Castillo:2025wyr,
    author = "Castillo, Daniel and Frycz, Dorian and Benavente, Beatriz and Men{\'e}ndez, Javier",
    title = "{Two-neutrino {\ensuremath{\beta}}{\ensuremath{\beta}} decay to excited states at next-to-leading order}",
    primaryClass = "nucl-th",
    doi = "10.1016/j.physletb.2026.140306",
    journal = "Phys. Lett. B",
    volume = "875",
    pages = "140306",
    year = "2026"
}

@misc{Horoi:2026fxp,
    author = "Horoi, Mihai and Neacsu, Andrei",
    title = "{}",
    eprint = "2605.21657",
    archivePrefix = "arXiv",
    primaryClass = "nucl-th",
    month = "5",
    year = "2026"
}

@article{Menendez:2011qq,
    author = "Menéndez, J. and Gazit, D. and Schwenk, A.",
    title = "{Chiral two-body currents in nuclei: Gamow-Teller transitions and neutrinoless double-beta decay}",
    primaryClass = "nucl-th",
    doi = "10.1103/PhysRevLett.107.062501",
    journal = "Phys. Rev. Lett.",
    volume = "107",
    pages = "062501",
    year = "2011"
}

@article{Wang:2019hjy,
    author = "Wang, X. B. and Hayes, A. C. and Carlson, J. and Dong, G. X. and Mereghetti, E. and Pastore, S. and Wiringa, R. B.",
    title = "{Comparison between Variational Monte Carlo and Shell Model Calculations of Neutrinoless Double Beta Decay Matrix Elements in Light Nuclei}",
    primaryClass = "nucl-th",
    reportNumber = "LA-UR-19-25587",
    doi = "10.1016/j.physletb.2019.134974",
    journal = "Phys. Lett. B",
    volume = "798",
    pages = "134974",
    year = "2019"
}

@article{Weiss:2021rig,
    author = "Weiss, Ronen and Soriano, Pablo and Lovato, Alessandro and Menendez, Jaview and Wiringa, R. B.",
    title = "{Neutrinoless double-{\ensuremath{\beta}} decay: Combining quantum Monte Carlo and the nuclear shell model with the generalized contact formalism}",
    primaryClass = "nucl-th",
    reportNumber = "LA-UR-21-32144",
    doi = "10.1103/PhysRevC.106.065501",
    journal = "Phys. Rev. C",
    volume = "106",
    number = "6",
    pages = "065501",
    year = "2022"
}

@article{Pastore:2017uwc,
    author = "Pastore, S. and Baroni, A. and Carlson, J. and Gandolfi, S. and Pieper, Steven C. and Schiavilla, R. and Wiringa, R. B.",
    title = "{Quantum Monte Carlo calculations of weak transitions in $A=6–10$ nuclei}",
    primaryClass = "nucl-th",
    doi = "10.1103/PhysRevC.97.022501",
    journal = "Phys. Rev. C",
    volume = "97",
    number = "2",
    pages = "022501",
    year = "2018"
}

@article{King:2020wmp,
    author = "King, G. B. and Andreoli, L. and Pastore, S. and Piarulli, M. and Schiavilla, R. and Wiringa, R. B. and Carlson, J. and Gandolfi, S.",
    title = "{Chiral Effective Field Theory Calculations of Weak Transitions in Light Nuclei}",
    primaryClass = "nucl-th",
    doi = "10.1103/PhysRevC.102.025501",
    journal = "Phys. Rev. C",
    volume = "102",
    number = "2",
    pages = "025501",
    year = "2020"
}

@article{Ney:2021anu,
    author = "Ney, E. M. and Engel, J. and Schunck, N.",
    title = "{Two-body weak currents in heavy nuclei}",
    primaryClass = "nucl-th",
    doi = "10.1103/PhysRevC.105.034349",
    journal = "Phys. Rev. C",
    volume = "105",
    number = "3",
    pages = "034349",
    year = "2022"
}

@article{Horiuchi:2021pup,
    author = "Horiuchi, Wataru and Sato, Toru and Uesaka, Yuichi and Yoshida, Kenichi",
    title = "{Electron wave functions in beta-decay formulas revisited (I): Gamow{\textendash}Teller and spin-dipole contributions to allowed and first-forbidden transitions}",
    primaryClass = "nucl-th",
    reportNumber = "J-PARC-TH-0240: KUNS-2860",
    doi = "10.1093/ptep/ptab069",
    journal = "PTEP",
    volume = "2021",
    number = "10",
    pages = "103D03",
    year = "2021",
    note = "[Erratum: PTEP 2022, 059301 (2022)]"
}

@article{Nitescu:2022tlh,
    author = "Ni{\c{t}}escu, O. and Stoica, S. and {\v{S}}imkovic, F.",
    title = "{Exchange correction for allowed {\ensuremath{\beta}} decay}",
    primaryClass = "physics.atom-ph",
    doi = "10.1103/PhysRevC.107.025501",
    journal = "Phys. Rev. C",
    volume = "107",
    number = "2",
    pages = "025501",
    year = "2023"
}

\end{document}